# Laser writing of coherent colour centres in diamond

Yu-Chen Chen[1], Patrick S. Salter[2], Sebastian Knauer[3], Laiyi Weng[1], Angelo C. Frangeskou[4], Colin J. Stephen[4], Philip R. Dolan[1], Sam Johnson[1], Ben L. Green[4], Gavin W. Morley[4], Mark E. Newton[4], John G. Rarity[3], Martin J. Booth[2] and Jason M. Smith[1,*]

[1] Department of Materials, University of Oxford, Parks Road, Oxford OX1 3PH, UK

[2] Department of Engineering Science, University of Oxford, Parks Road, Oxford OX1 3PJ, UK

[3] Department of Electronics and Electrical Engineering, University of Bristol, Merchant Venturers Building, Woodland Road, Bristol, BS8 1UB, UK

[4] Department of Physics, University of Warwick, Coventry, CV4 7AL, UK

*Email: jason.smith@materials.ox.ac.uk

Optically active point defects in crystals have gained widespread attention as photonic systems that can find use in quantum information technologies [1,2]. However challenges remain in the placing of individual defects at desired locations, an essential element of device fabrication. Here we report the controlled generation of single nitrogen-vacancy (NV) centres in diamond using laser writing [3]. The use of aberration correction in the writing optics allows precise positioning of vacancies within the diamond crystal, and subsequent annealing produces single NV centres with up to 45% success probability, within about 200 nm of the desired position. Selected NV centres fabricated by this method display stable, coherent optical transitions at cryogenic temperatures, a pre-requisite for the creation of distributed quantum networks of solid-state qubits. The results illustrate the potential of laser writing as a new tool for defect engineering in quantum technologies.

The NV centre is one of an increasing number of point defects in wide band-gap materials such as diamond and silicon carbide that show promise as long-lived quantum light sources and provide an optical interface with coherent electronic and nuclear spins [4-12]. Realization of entangled networks for sensing or distributed quantum computing with these systems [13, 14] also requires coherence of the optical transition, which some defects have displayed at cryogenic temperatures [7, 15]. In the negatively charged NV centre this coherent interface has been used to demonstrate spin-photon entanglement [16] and entanglement between distant spins [17].

To realise technological applications it will be necessary to place coherent colour centres at desired locations in a crystal so that they can be integrated with optical and electronic components, on length scales of 10 nm – 1μm [18-20]. This presents a significant challenge, since most commonly used colour centre fabrication techniques, such as ion implantation and electron irradiation, create significant residual damage to the crystal lattice, which degrades the colour centre properties via effects of strain and the creation of additional unwanted defects in the local environment. Despite some recent progress in this area [21, 22], new methods that achieve accurate placement of defects with minimal residual lattice damage are of significant interest.

Direct laser writing with ultrafast pulsed lasers has been shown to be a powerful tool for fabricating photonic circuits in glass to implement quantum optical processing [23, 24]. Here we show that it is equally powerful when used to write vacancies directly into a crystal as a starting point for colour centre formation. This method has a number of attractive features: the highly nonlinear process of laser writing [3] results in localised vacancy generation without any damage to the overlaying material; the inclusion of adaptive optics for aberration correction [25] allows small focal volumes (< 0.1 µm$^3$) at any depth within the crystal; and the laser pulse energy can be tuned with high precision to control the number of vacancies generated. We show here that after applying this process to diamond samples in which nitrogen atoms are present at low concentrations, subsequent annealing produces single, high quality NV centres 'on-demand'.

The diamond samples used were commercially available single crystals grown by chemical vapour deposition, with 'background' nitrogen density below <1000 atoms per µm$^3$. Regions were chosen for processing in which no 'native' NV centres were present. Single laser writing pulses of wavelength 790 nm and duration 300 fs were delivered to each site in 25 × 20 square grids with a separation of 5 µm at a depth of 50 µm. Along one axis of the grids, the pulse energy $E_p$ was varied between 16.0 nJ and 61.8 nJ in 24 increments to generate varying degrees of damage to the lattice. Along the other axis, 20 identical pulses were delivered to facilitate statistical analysis of the results for each pulse energy.

Figure 1a shows a photoluminescence (PL) image of one of the samples (sample A) immediately after laser writing. $E_p$ increases from the bottom to the top of the image. Visible PL emission was produced from sites that had been exposed to pulses with $E_p > 31$ nJ, and inspection of the PL spectra (Figure 1b, upper trace) confirmed the presence of photo-generated vacancies. This threshold energy for visible vacancy fluorescence, $E_1$, is indicated by a red line on the image. Figure 1c shows the PL image of the same array after annealing (see Methods). Fluorescence was then observed at several sites for which $E_p < E_1$ indicating the formation of new colour centres. Some sites in rows for which $E_p > E_1$ no longer showed fluorescence, suggesting that the majority of the isolated vacancies in these rows had been removed as a result of the anneal step [26]. Fluorescence from sites up to $E_2 = 36.4$ nJ (the green line in figure 1c) was exclusively from NV⁻ centres, with characteristic zero phonon line at 637 nm and broad phonon sideband (figure 1b, lower spectrum). Features created with $E_p > E_2$ showed additional broad-band fluorescence, suggesting that the damage created at these sites exceeded the threshold for graphitization during the anneal [27].

Hanbury Brown and Twiss (HBT) measurements of the two-photon correlation function $g^{(2)}(\delta t)$ were carried out for each of the sites with $E_p < E_1$ to determine the number of NV centres present. A typical dataset recorded from a site with a single NV centre is shown in figure 2a, revealing a characteristic dip well below 0.5 at $\delta t = 0$. No background correction has been applied to the data. A histogram of the $g^2(0)$ values from all sites measured is shown in figure 2b. Two clear populations emerge, one with $g^2(0) < 0.32$ and another with $0.32 < g^2(0) < 0.65$, which we attribute to the presence of one and two NV centres respectively. A few sites showed $g^2(0)$ between 0.65 and 0.9, which we attribute to the presence of three NV centres. Figure 2c shows a spatial map of the locations of the sites with single, double, and triple NV centres, and figure 2d shows the row statistics as functions of laser pulse energy. The probability of generating a single NV at a lattice site was found to be as high as 45%, at $E_p = 25.7$ nJ. The total number of NV centres generated per row (black data points) reveals a clear general trend for more NV centres to be generated at higher pulse energies, but with a region of significant deviation from this trend spanning five rows, most likely due to non-uniform nitrogen distribution in the sample.

From the PL image the location of each NV centre in the image plane can be determined to within < 100 nm. Figure 2e shows a magnified image of a section of the NV array with superimposed grid corresponding to the target positions. A histogram of the measured displacements of the NV defects from the target points is shown in figure 2f. The solid line is a fit of the 2D distribution function $f(r) = Are^{-r^2/r_0^2}$ where $r$ is the radial displacement and $A$ and $r_0$ are fitting parameters. $r_0$ was found to be 196 ± 20 nm, consistent with the thermal diffusion length for vacancies during the anneal process (see Supplementary information).

Photoluminescence excitation (PLE) spectroscopy (see methods) was used to probe the zero-phonon lines of the NV centres at $T = 4.2$ K, and reveals examples of colour centres with highly coherent transitions. Figure 3a shows a selection of PLE scans from three different NV centres in sample B, showing line widths of 13.5 ± 0.3 MHz, 12.0 ± 0.7 MHz, and 27.5 ± 1.2 MHz. The two narrowest lines are consistent with the Fourier transform limit of 12.4 ± 0.1 MHz for an NV centre in bulk diamond based on a fluorescence relaxation time of 12.8 ± 0.1 ns (see Supplementary information). Of 34 NV centres measured in this sample, seven revealed line widths below 30 MHz and four under 20 MHz. Repeated scans of one of these NV centres (Figure 3b) revealed that the transition remained stable for a prolonged period with small fluctuations in its peak position producing an inhomogeneously broadened line of width 16.1 MHz. After prolonged excitation the NV ionized and an optical repump was required to restore the negative charge state. The repump pulse (λ=532 nm) ionized other defects in the vicinity of the NV centre, changing the local electric field and shifting its transition energy by about 70 MHz via the Stark effect (Fig. 3c). Such shifts can be suppressed by using a repump wavelength tuned to 575 nm [21,28], or corrected using dynamic stabilization techniques [29].

Finally, Hahn echo measurements were carried out to measure the spin coherence times of a selection of defects. Figure 4a shows an example echo decay signal recorded from an NV centre created with a low laser pulse energy, $E_p$ = 19.6 nJ, and displays $T_2$ = 48 ± 2 μs. Of the seven NV centres measured, most showed $T_2$ times between 30 and 80 μs, while one showed a spin echo that remained strong beyond the 100 μs time duration of our apparatus. These values are fairly typical for NV centres in diamond containing a naturally occurring concentration of $^{13}$C nuclei [30]. No particular correlation was observed between optical and spin coherence in our study, although it is worth noting that the NV centre with the longest $T_2$ time also showed a narrow optical line width of 25 MHz.

In summary, our results show that femtosecond laser processing can be used to 'write' vacancies into diamond in 3D with the control required to form optically coherent NV centres at desired locations. The technique is quite general and may be adapted for the generation of other colour centres (most of which involve vacancies) or point defects in other wide band gap materials. Our results show sufficient positioning accuracy for placing NV centres in a variety of optical structures including waveguides or whispering gallery resonators or under solid immersion lenses, and for near-deterministic positioning in nano-beams and photonic crystal cavities. The accuracy is currently limited by the diffusion of vacancies during annealing, and further improvement may be achieved by modifying the background nitrogen concentration and adjusting the annealing recipe.

**Methods**

The samples used were 'electronic grade' single crystal plates from Element Six Ltd, with (001) crystal orientation. Laser writing was performed using a regeneratively amplified Ti:Sapphire laser. For aberration correction a liquid crystal phase-only spatial light modulator was imaged onto the back aperture of a 60× 1.4NA oil immersion objective and adjusted to minimise the pulse energy needed to produce visible fluorescence during processing. Pulse energies were controlled using a rotatable half-wave plate in conjunction with a Glan-Laser polariser and measured before the microscope objective lens. Annealing was carried out for 3 hours at 1000 degrees Celsius under pure nitrogen gas. PL imaging was carried out using a home-made scanning confocal microscope and spectroscopy with a 500 mm spectrograph. NV locations were measured as the centre of a 2D Gaussian function fitted to the PL image. For PLE measurements, a tunable external cavity diode laser with line width 1 MHz was used for excitation, its frequency tuned through the ZPL resonance, and the phonon sideband emission intensity monitored. For the Hahn echo measurements a dc magnetic field of 6.7 mT was used to isolate the $m_s = 0 \leftrightarrow m_s = -1$ transition in the ground state spin manifold. The amplitude of the fluorescent signal was then measured as a function of the microwave pulse separation. The duration of a π/2-pulse was 20 ns (40 ns for a π-pulse) such that the pulse bandwidth was larger than the splitting, allowing full excitation of the ESR -1 electron transition. The signal is normalised to the equilibrium PL intensity between the -1 and 0 state of the optical pulse. The Hahn echo decay is measured over the first decay and on the revival. Further details of experimental methods are provided in the online supplementary information.

**Acknowledgements**

Y-C Chen would like to thank DeBeers for financial support. This work was supported by grants from the European Commission (WASPS project, grant agreement no 618078), the UK Engineering and Physical Sciences Research Council, (EP/M013243/1) and The Leverhulme Trust.

Figures:

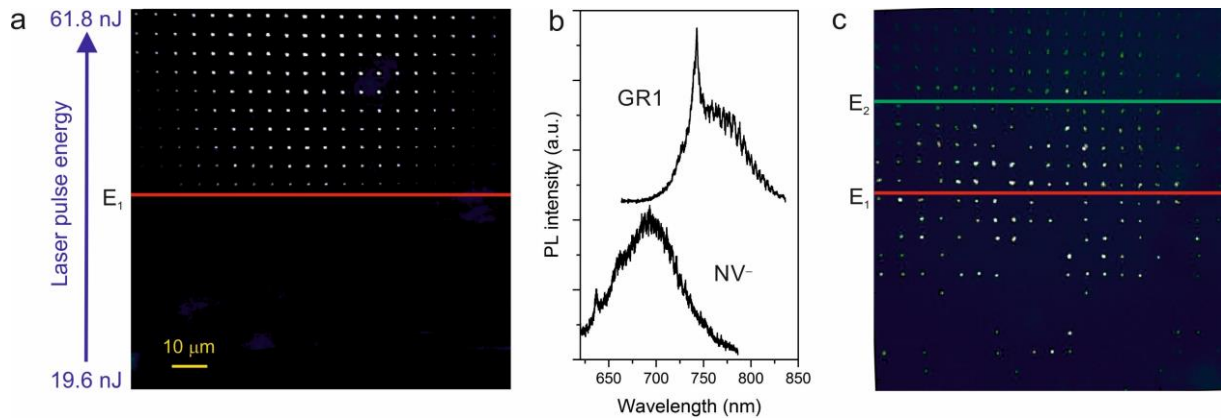

**Figure 1 | NV centre generation using laser writing**. **a**, Photoluminescence image of the 25 x 20 array immediately after laser processing (before annealing). The laser pulse energy increases from the bottom to the top of the image. The red line at pulse energy $E_1$ indicates the lowest energy laser pulse that produces visible fluorescence. The drop off in intensity of features toward the edge of the array is due to field aberrations in the PL microscope. **b**, Typical spectra measured from points in fig 1a (upper plot), characteristic of GR1 (single vacancy) defects and from figure 1c (lower plot) characteristic of the negatively charged nitrogen-vacancy (NV⁻) centre. **c,** PL image of the same region of sample after the annealing process, showing NV emission from multiple sites processed with pulse energies below *$E_1$*. The green line at pulse energy $E_2$ indicates the graphitization threshold.

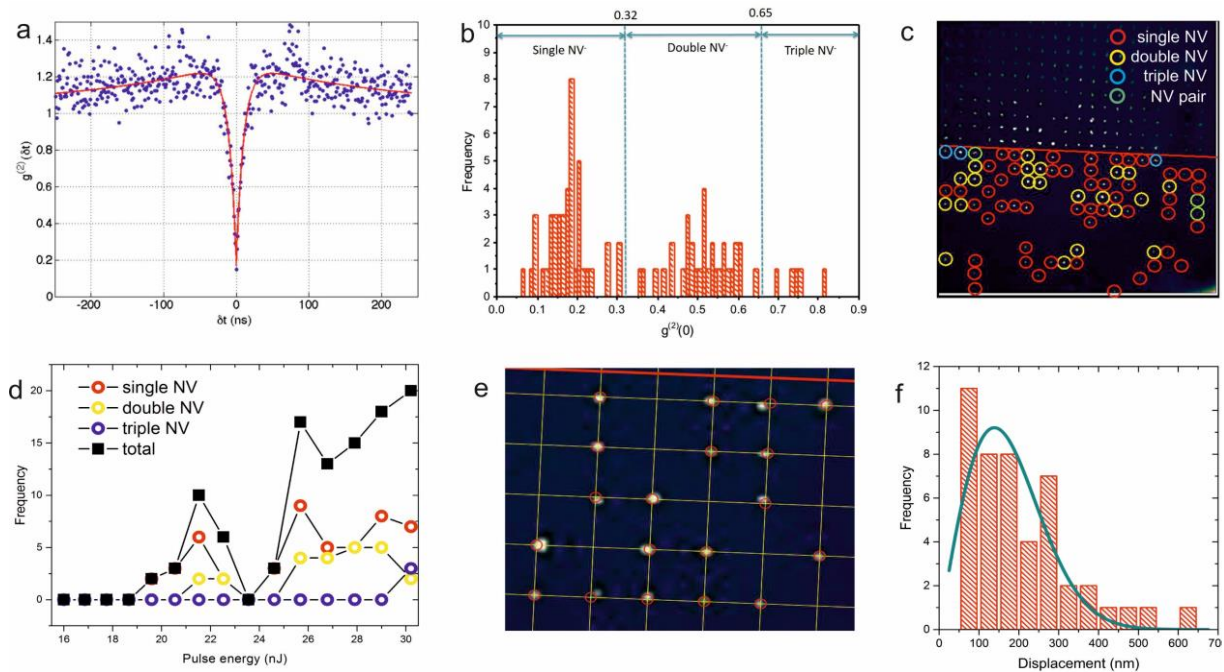

**Figure 2 | Statistics and positioning accuracy of NV generation using laser writing**. **a**, Histogram showing the two-photon correlation function $g^{(2)}(\delta t)$ from a single NV centre; **b**, Histogram of $g^{(2)}(0)$ for the different laser processing sites, allowing identification of sites of single, double, and triple NV centre generation. **c**, Map of the number of NV centres generated at different sites. 'NV pair' refers to a double NV where the two defects are spatially resolved. **d**, Histogram of the number of single (red), double or 'pair' (yellow), and triple (blue) NV's generated in each row of 20 sites as a function of laser pulse energy measured before the objective lens in the writing apparatus. The total number generated per row is shown in black. **e**, Magnified image of NV centre fluorescence relative to the laser processing grid. Red circles centred on the grid points are 1 μm in diameter. **f**, Histogram of the displacement in the image plane for the single NV centres measured, fitted with a cylindrical distribution function (see text).

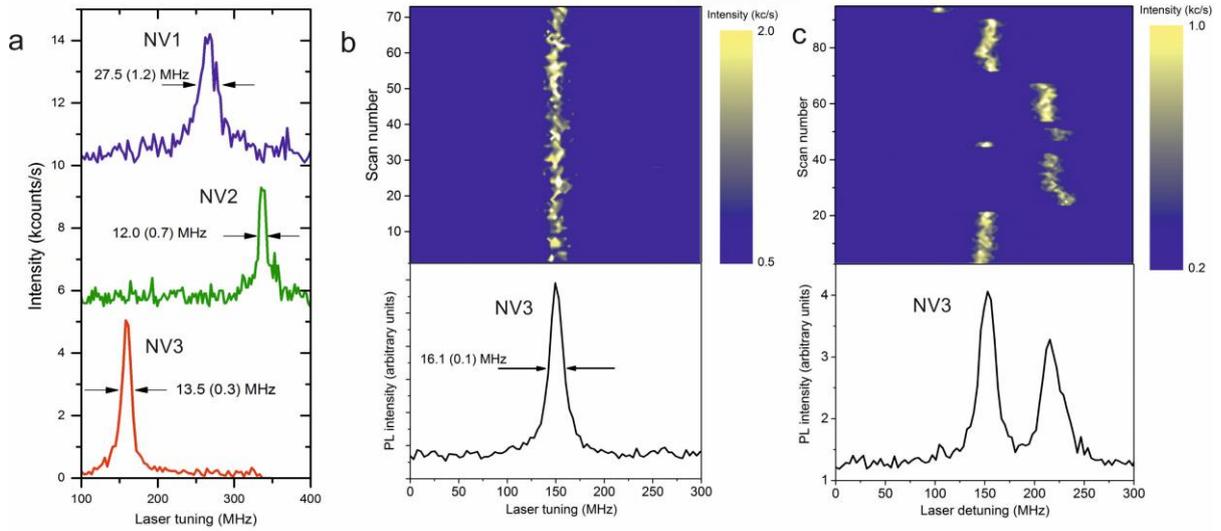

**Figure 3 | Spectral properties of single laser-written NV centres at 4.2 K**. **a**, Photoluminescence excitation (single sweep) of three different NV centres, with two showing Lorentzian peaks below 14 MHz in width. Full-width-at-half-maximum values from Lornetzian peak fits are give, with errors in parentheses; **b**, Colourscale map of repeated PLE spectra of NV3, showing a stable line over 70 laser sweeps with an inhomogeneous line width of 16.1 MHz. **c**, Spectral jumping as a result of a 532 nm re-pump pulse required to restore the negative charge state upon ionization. Lower plots in b and c are aggregates of the consecutive sweeps in the colourscale images.

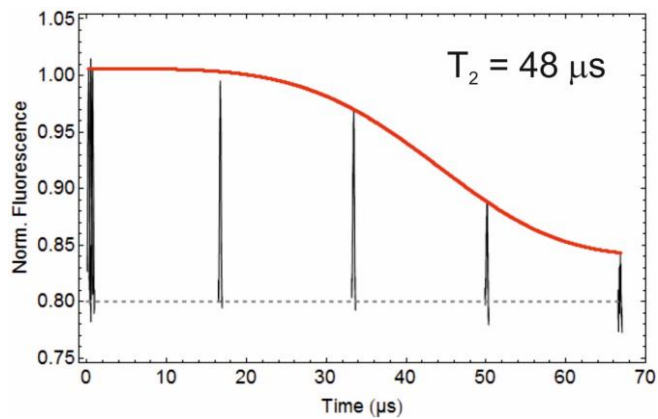

**Figure 4 | Spin resonance properties of laser-written NV centres at 300K**. Hahn echo data for an NV centre created with pulse energy 19.6 nJ, fitted with the function $I(\tau) = y_1 e^{-(\tau/T_2)^n} + y_0$, where the exponent *n* is a free parameter and $y_0$, $y_1$ and $T_2$ are fitting parameters.

# Supplementary Information

## A. Supplementary data

**Samples**

The paper contains data recorded from two separate samples, labelled A and B, both of which were commercial 'electronic grade' single crystal plates from Element Six as described in the Methods section, and were processed in an identical manner. The PL images and spectra in figure 1, NV population data in figures 2a-e, and spin coherence data in figure 4 were recorded from sample A. NV positioning data in figure 2f and PLE data in figure 3 were recorded from sample B.

Sample B showed about half the probability of generating NV centres compared with sample A for the same laser pulse energy. We attribute this reduced NV generation rate to a lower nitrogen concentration (the nitrogen concentrations in both samples are below the limit at which bulk electron Paramagnetic Resonance measurements can measure them.)

**Additional PL spectra of laser processed sites**

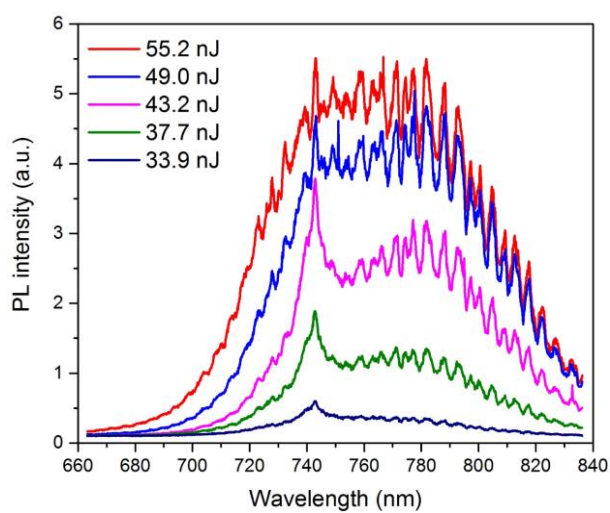

Figure S1 | Fluorescence spectra from laser processed sites (pre-anneal) at different pump pulse energies (sample A).

Fluorescence spectra were recorded from multiple sites in sample A prior to annealing (main text figure 1a). Figure S1 shows recorded spectra as a function of the laser writing pulse energy. These spectra reveal that the GR1 peak at 740 nm is visible for $E_p > 30$ nJ but at higher pulse energies becomes swamped by a broad signal at longer wavelengths originating from extended defects. The rapid modulation in the signal for wavelengths above 750 nm is due to an etaloning effect in the CCD camera.

Figure S2 shows typical fluorescence spectra recorded for $E_p > E_1$ in figure 1c, in which NV-emission is seen along with additional broad-band fluorescence suggestive of a high density of extended defects remaining in the material. Wherever combined NV and BB fluorescence was observed the photon autocorrelation dip was found to be significantly suppressed.

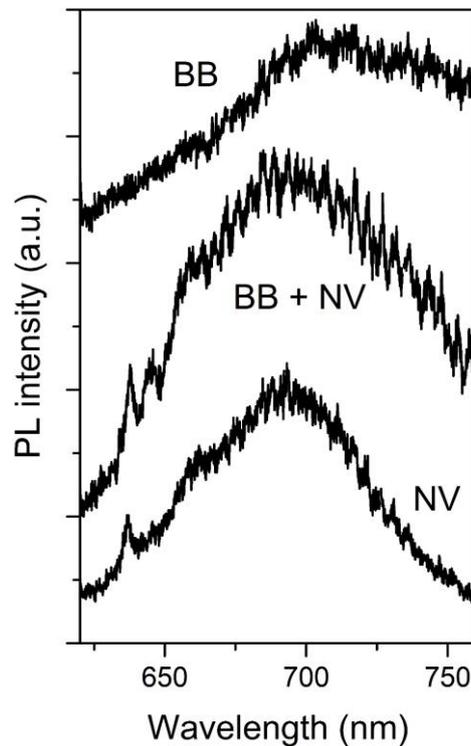

Figure S2 | Additional PL spectra recorded from a post-anneal sample at sites which were processed with laser pulse energies greater than $E_2$. Broad-band (BB) emission is observed (top plot), and spectra in which BB and NV emission are both present (middle plot). The spectra are offset in the vertical direction for clarity.

**Time-resolved Photoluminescence of NV3 in Sample B**

The excited state lifetime NV3 in sample B (one of the defects showing a narrow PLE line width at a temperature of 4.2 K) was measured directly from the time-resolved photoluminescence decay (TRPL) using the time-correlated single photon counting (TCSPC) technique. Excitation was with a 532 nm frequency-doubled diode-pumped-solid-state laser with 50 ps pulse durations (PicoQuant LDH-FA), and detection with an Excelitas SPCM AQR-14 single photon counting detector. Timing was recorded with an Edinburgh Instruments TCC900 PCI card. The decay function was recorded by measuring the TCSPC signal when focused on the NV defect then substracting the TCSPC signal from a nearby region of bare diamond. The instrumental response function was also measured by removing the laser blocking filter and inserting a neutral density filter to measure attenuated reflection of the laser. The instrumental response function was then convoluted with a single exponential decay and fitted to the NV centre decay function, with a lifetime $T_1$ and amplitude introduced as fitting parameters. Figure S3 shows a plot of the decay function after background removal, and the best fit to the data for a single exponential decay of lifetime $T_1$ convoluted with the instrumental response.

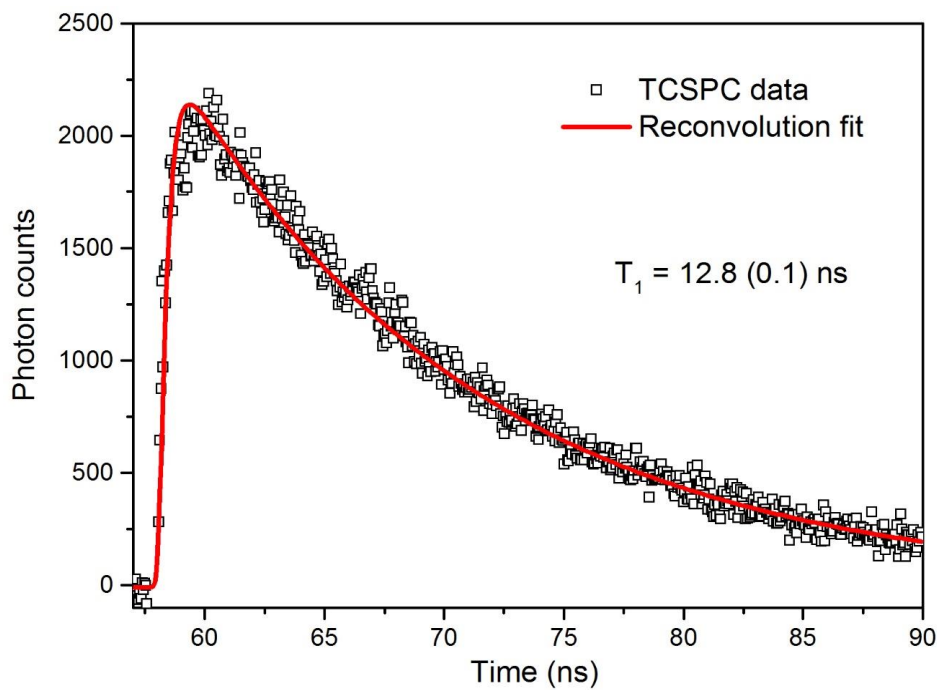

Figure S3 | Time-resolved photoluminescence of NV3 in sample B at 4.2K, fitted with a single exponential decay convoluted with the instrumental response of the TCSPC apparatus.

## B. Details of experimental methods

**Laser writing**

Direct Laser Writing (DLW) was carried out using a regeneratively amplified Ti:Sapphire laser (Spectra Physics Solstice). The optical layout for the laser processing is shown in Figure S1a.

The laser beam was expanded onto a liquid crystal phase-only spatial light modulator (SLM) (Hamamatsu X10468-02), which was imaged in a 4*f* configuration onto the back aperture of a 60× 1.4NA Olympus PlanApo oil immersion objective. The diamond sample was mounted on precision translation stages (Aerotech x-y: ABL10100; z: ANT95-3-V) providing three dimensional control. An LED illuminated transmission microscope provided visualisation of the sample during processing. Prior to the objective the laser pulse was linearly polarised and had a duration which was measured to be 250 fs using an intensity autocorrelator (APE Pulsecheck). The pulse energy at focus will be slightly increased due to dispersion in the objective lens.

The use of high numerical aperture focussing is necessary to limit the size of the laser focus, and move into a regime appropriate for the generation of single localised NVs as opposed to NV ensembles that tend to form around regions of laser induced graphitisation [Pimenov]. However, using high NA optics to focus beneath the surface of diamond introduces significant amounts of spherical aberration, due to refraction at the oil/diamond interface where there is a large step in refractive index. It is necessary to correct the aberration to restore optimum optical performance, which has also been shown to enhance the light matter interaction when laser writing sub-surface graphitic tracks [Sun]. Thus, the phase pattern displayed on the SLM was controlled to remove any system and specimen induced aberrations by experimentally minimising the laser pulse energy required for visible modification of the diamond lattice. The effect of the aberration correction is to provide dramatic reduction in size of the laser focus (Fig 1b). Since the spherical aberration at the diamond/oil interface is the dominant aberration source, the correction produces a particularly marked improvement in the axial (z) direction. We have shown previously that the use of adaptive optics enables accurate laser fabrication in diamond to depths of at least 220 µm [Simmonds].

The focal spot of the laser in the diamond is expected to be close to the diffraction-limit, with transverse and axial dimensions (full width half maximum) of 350 nm and 2 µm respectively, giving a focal volume of around 0.12 µm$^3$. However, the volume over which the vacancies are generated is likely to be considerably smaller, since the interaction of the pulse with the diamond is highly non-linear [Lagomarsino] and there is expected to be a distinct threshold for optical breakdown which may be exploited [Joglekar, Juodkazis].

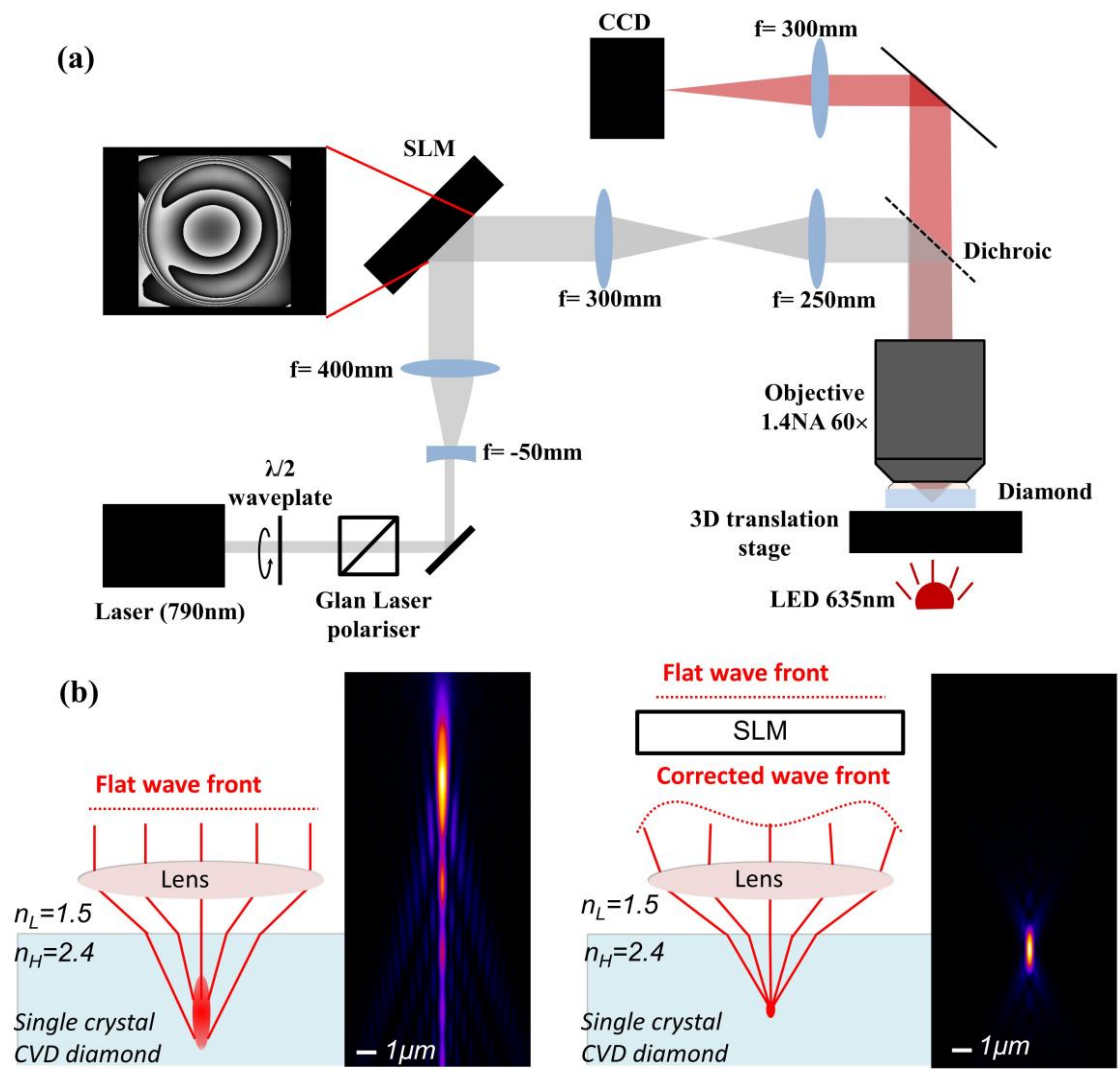

Figure S4 | Aberration-corrected laser writing. **a**, Schematic of laser processing apparatus. The phase pattern displayed on the SLM compensated optical aberrations in the system and those introduced by focussing into the diamond. **b**, The effect of aberration correction on the laser focal intensity distribution when focused through the planar oil/diamond interface to a depth of 50 µm. Without correction the focus is drastically elongated in z by spherical aberrations. With aberration correction good z resolution is achieved.

**Pulse energy control**

The laser pulse energy was precisely controlled using a rotatable half waveplate before a Glan-Laser polariser. Pulse energies for the 25 waveplate settings used for the fabrication array are shown in table S1 below. Pulse energy measurements (2nd column) were taken immediately before the objective. The final column gives the pulse energy reduced by a factor of 0.7 due to the transmission loss of the objective lens, based on the manufacturers specifications. For the pulse energy at the focus, one must also take into account Fresnel reflection at the diamond interface which can be significant due to the large angles involved.

| Angle of $\lambda/2$ waveplate (degrees) | Pulse energy before objective (nJ) | Pulse energy (nJ) after objective |
|---|---|---|
| 5 | 118.0 | 82.6 |
| *30μm gap* | | |
| 3.6 | 61.8 | 43.2 |
| 3.4 | 55.2 | 38.6 |
| 3.2 | 49.0 | 34.3 |
| 3 | 43.2 | 30.2 |
| 2.8 | 37.7 | 26.4 |
| 2.75 | 36.4 | 25.5 |
| 2.7 | 35.1 | 24.6 |
| 2.65 | 33.9 | 23.7 |
| 2.6 | 32.6 | 22.8 |
| 2.55 | 31.4 | 22.0 |
| 2.5 | 30.2 | 21.1 |
| 2.45 | 29.0 | 20.3 |
| 2.4 | 27.9 | 19.5 |
| 2.35 | 26.8 | 18.7 |
| 2.3 | 25.7 | 18.0 |
| 2.25 | 24.6 | 17.2 |
| 2.2 | 23.6 | 16.5 |
| 2.15 | 22.5 | 15.8 |
| 2.1 | 21.5 | 15.1 |
| 2.05 | 20.5 | 14.4 |
| 2 | 19.6 | 13.7 |
| 1.95 | 18.7 | 13.1 |
| 1.9 | 17.7 | 12.4 |
| 1.85 | 16.9 | 11.8 |
| 1.8 | 16.0 | 11.2 |

Table S1 Laser pulse energies used for the writing of vacancies. The energies measured before the objective are the ones quoted in the main text.

The first row of the fabrication grid was at sufficient pulse energy (118 nJ) for considerable graphitisation and could hence be seen using a standard transmission microscope. These were used as location markers since the rest of the fabrication array produced too little structural modification within the diamond to be seen by standard widefield microscopy methods and were only revealed via their photoluminesence.

**Annealing**

Annealing was carried out in a tube furnace (Elite Thermal Systems TSH16/50/180-2416). The diamonds were placed in an alumina boat and buried in a sacrificial diamond grit (Element Six Micron+). The furnace was purged with dry nitrogen boil-off to minimize oxidization and graphitization of the diamond surface.

Three different annealing recipes were tried, at 900, 1000 and 1200 degrees Celsius for three hours, and a multi-step process following Chu et al [21] involving 4 hours at 400 degrees Celsius, 2 hours at 800 degrees Celsius and 2 hours at 1200 degrees Celsius, with 3 degrees Celsius / minute ramping rate between steps.

The sample annealed at 900 degrees we observed that a few NV$^-$ centres were generated on the original damage position, but that most of the GR1 defects and other damage remained. This result is consistent with literature reports that most extended vacancy defects created by implantation process are destroyed at temperatures above about 1000 degrees Celsius as reported in reference [26] of the main text. The 1000 degrees anneal produced many single NV defects as reported in the main text. Subsequent annealing of the same sample at 1200 degrees Celsius for 24 hours resulted in no new NV$^-$ centres, and all the generated NV$^-$ centres in previous array disappeared. No NV centre was generated after the multi-step process annealing from reference [21] in the main text. We concluded that a 3 hours 1000 degrees Celsius annealing was the optimal condition for this laser fabrication method.

During annealing, the vacancies are expected to diffuse an average distance $\sqrt{Dt}$ where $D$ is the diffusivity and $t$ is the anneal time. For an initial vacancy distribution that is point-like in the $xy$ plane and elongated in the $z$ direction, isotropic diffusion yields a cylindrically symmetric vacancy distribution after annealing, following

$$n_V(r) = Are^{-r^2/4Dt} \tag{1}$$

The fit shown in figure 2f of the Letter therefore provides $\sqrt{Dt} = r_0/2 = 98 \pm 1$ nm, giving a value of $D = 3.7 \times 10^{-14}$ cm$^2$ s$^{-1}$. Diffusion of vacancies in diamond is a thermally activated hopping process the activation energy for which can be expressed as

$$\Delta = -kT\ln\left(\frac{D}{D_0}\right), \qquad (2)$$

where $kT$ is the thermal energy. Using $D_0 = 3.6 \times 10^{-6}$ cm$^2$ s$^{-1}$ [Hu] we find that $\Delta = 2.0$ eV, consistent with values presented in other reports [Bernholc, Mainwood, Orwa].

**Photoluminescence measurements and positioning accuracy**

Photoluminescence (PL) imaging and the spectroscopy were performed using a custom-built scanning confocal microscope, the full design and specifications of which are reported in reference [Grazioso]. Excitation was performed using a frequency-double YAG laser (λ=532 nm) with a maximum power delivery to the sample of 4 mW. Spectra were recorded with an

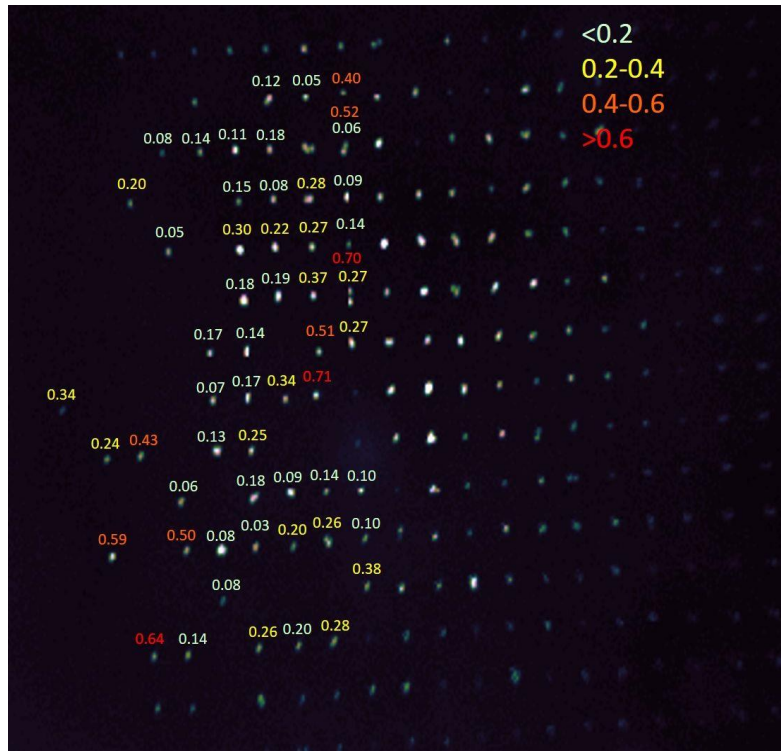

Figure S5 | Positioning accuracy of NV generation on sample B. Labels show the displacement of NV centres from their respective grid positions (in micrometres), after removal of a quadratic field distortion. The labels are colour-coded into four magnitude groups to reveal any systematic trends that would indicate further uncompensated field distortion.

Acton SpectraPro 0.75 m monochromator fitted with a back-illuminated CCD camera (Princeton Spec-10 100B), with > 90% quantum efficiency across the wavelength range of interest. A 532 nm laser clean-up filter is used in excitation, combined with a 540 nm dichroic beam splitter and a 532 nm blocking notch filter in the collection optics. When recording a PL image, a 650 long-pass filter is inserted in the fluorescence collection path to block the diamond Raman emissions (all filters were from Semrock).

The positioning accuracy was determined from the PL images by fitting a 2D Gaussian surface to the image of each single NV, and measuring its displacement relative to the uniform grid pattern. The spatial resolution of the microscope (~500 nm) combined with the intensity of the single NV fluorescence images (~10,000 counts) allows Gaussian fitting with a standard error in the position of the NV centres of between 50 and 100 nm.

After subtracting a quadratic field distortion for the PL microscope, determined using a pre-annealed array, these positions were compared with the uniform grid pattern locations targeted in the processing microscope. Figure S5 shows a map of the resultant displacements. That there is little evidence of systematic order in the displacements suggests that the major field distortions have been corrected, and that the remaining displacements reflect the random location of the NV centres about the target sites.

**Hanbury Brown and Twiss measurements and data fitting**

Hanbury Brown and Twiss measurements of the laser-generated NV$^-$ centres were recorded with continuous wave 532 nm excitation at a power of 0.74 mW corresponding to 0.87P$_{sat}$. The fluorescence was spectrally filtered using a 650 nm long pass filter.

The autocorrelation function results were fitted by the standard three-level system model of the autocorrelation function with a fixed background

$$g^{(2)}(\delta t) = g^{(2)}(0) + 1 - c * \exp\left(\frac{-|\delta t|}{\tau_2}\right) + (c - 1) * \exp\left(\frac{-|\delta t|}{\tau_3}\right) \qquad (3)$$

where c, $\tau_2$ and $\tau_3$ are related to the inter-level rate constants as described in [Kurtsiefer]. The solid red line in Fig. 2(e) is least-squares fits of equation (1) to the measured data.

**Photoluminescence excitation measurements**

For PLE measurements the sample was cooled to 4.2 K in a liquid helium bath cryostat with built-in confocal microscope [Grazioso]. An external cavity diode laser (Toptica DL100) was scanned through resonance with the NV- zero phonon line at a wavelength of 637 nm, and the wavelength measured directly using a wavemeter (High Finesse WSU-30). No microwave modulation was applied, so the spin population time of the NV centres is very long, and so only the highly cycling upper branch $m_s = 0$ transition is observed.

Repump pulses were delivered from a frequency doubled diode-pumped solid state laser at 532 nm (CNI MGL-III-532), gated into 200 ms pulses using an acousto-optic modulator.

**Spin resonance measurements**

For the Hahn echo measurements ($\pi/2$-$\tau_0$-$\pi$-$\tau_1$-echo) we used a home-built scanning confocal microscope. For all the measurements an external magnetic field of 6.7 mT was applied and perfectly aligned in the perpendicular plate to the NV centre axis with pulsed and cw ODMR techniques in order to separate the -1 and +1 electron transitions. The splitting between the -1 and +1 fine transitions is big enough, ensuring the individual driving of the -1 transition. The amplitude of the fluorescent signal is measured as a function of the pulse separation. The time of the $\pi/2$-pulse is chosen to 20ns (40ns for $\pi$-pulse) (determined by Rabi oscillation measurements on the -1 transition). Therefore the bandwidth was larger than the splitting, allowing full excitation of the -1 electron transition. A single Hahn echo measurement, i.e. iteration of $\tau_1$ around the set value of $\tau_0$, consists of a few million iterations, leading to about 3500 counts for the optical pulses (assuming Poisson error). The signal is normalised to the equilibrium between the -1 and 0 state of the optical pulse. The Hahn echo decay is measured over the first decay and on the revival.

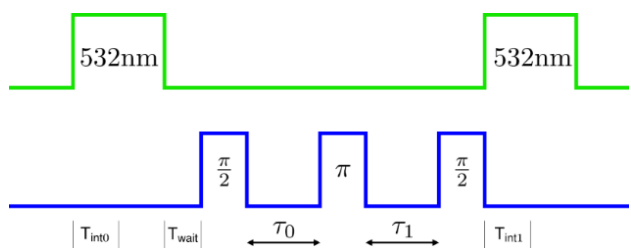

Figure S6 | Pulse sequence for the Hahn echo measurements. Laser gating in indicated in green and microwave gating in blue.